\documentclass[runningheads]{llncs}
\usepackage[T1]{fontenc}
\usepackage{graphicx}
\usepackage{booktabs}
\usepackage[misc]{ifsym}

\usepackage{amsmath,amsfonts,amssymb}
\usepackage{algorithm}
\usepackage{algpseudocode}
\usepackage{array}
\usepackage[caption=false,font=normalsize,labelfont=sf,textfont=sf]{subfig}
\usepackage{textcomp}
\usepackage{stfloats}
\usepackage{url}
\usepackage{verbatim}
\usepackage{graphicx}
\usepackage{cite}
\usepackage{bm}
\usepackage{siunitx}
\usepackage{orcidlink}
\usepackage[noabbrev,nameinlink]{cleveref}
\usepackage{multirow}

\hypersetup{hidelinks}

\title{Triggering hallucinations in model-based MRI reconstruction via adversarial perturbations}
\titlerunning{Triggering hallucinations in MRI reconstruction}

\authorrunning{Bu\u{g}day et al.}
\author{Suna Bu\u{g}day\inst{1} \and Yvan Saeys\inst{2,3} \and Jonathan Peck\inst{2,3}\thanks{Corresponding author: \texttt{Jonathan.Peck@UGent.be}}}

\institute{Faculty of Engineering and Architecture/Computer Engineering, \.{I}zmir Bak{\i}r\c{c}ay University, \.{I}zmir, Turkey \and
Department of Computer Science, Mathematics and Statistics, Ghent University, Ghent, Belgium \and
Data Mining and Modeling for Biomedicine Group, VIB Inflammation Research Center, Ghent, Belgium}

\newcommand{\comp}{\mathbb C}
\newcommand{\real}{\mathbb R}
\newcommand{\norm}[1]{\left\| #1 \right\|}
\newcommand{\reg}{\mathcal R}
\newcommand{\tv}{\mathrm{TV}}
\newcommand{\LComment}[1]{\Statex \hspace{\algorithmicindent} // #1}

\newcommand{\codeurl}{\url{https://github.com/saeyslab/adversarial-mri}}

\DeclareMathOperator{\ZF}{ZF}

\begin{document}

\maketitle

\begin{abstract}
Generative models are increasingly used to improve the quality of medical imaging, such as reconstruction of magnetic resonance images and computed tomography.
However, it is well-known that such models are susceptible to \emph{hallucinations}:
they may insert features into the reconstructed image which are not actually present in the original image.
In a medical setting, such hallucinations may endanger patient health as they can lead to incorrect diagnoses.
In this work, we aim to quantify the extent to which state-of-the-art generative models suffer from hallucinations in the context of magnetic resonance image reconstruction.
Specifically, we craft adversarial perturbations resembling random noise for the unprocessed input images which induce hallucinations when reconstructed using a generative model.
We perform this evaluation on the brain and knee images from the fastMRI data set using UNet and end-to-end VarNet architectures to reconstruct the images.
Our results show that these models are highly susceptible to small perturbations and can be easily coaxed into producing hallucinations.
This fragility may partially explain why hallucinations occur in the first place and suggests that a carefully constructed adversarial training routine may reduce their prevalence.
Moreover, these hallucinations cannot be reliably detected using traditional image quality metrics.
Novel approaches will therefore need to be developed to detect when hallucinations have occurred.

\end{abstract}

\section{Introduction}
When patients go in for a medical scan, such as magnetic resonance imaging (MRI) or computed tomography (CT), the resulting images must first be cleaned up or \emph{reconstructed} before they can be used for diagnostic purposes.
There are multiple reasons for this: patients never stay completely still during scans, which causes blurring; the physical measurement process is subject to thermal noise and electromagnetic interference; etc~\cite{bellon1986mr}.
Scanners also frequently undersample or \emph{accelerate} their measurements, which can lead to aliasing and other artifacts due to incomplete data~\cite{heckel2024deep}.
For MRI, acceleration is usually performed to save costs and to increase patient comfort: a fully-sampled MRI would take several hours compared to the few minutes patients typically spend inside such a scanner in practice~\cite{ballinger2013acquisition}.
For CT, minimizing the acquisition time is actually a necessity due to the exposure to a radioactive isotope which can damage tissues over an extended period of time.

\begin{figure}
    \subfloat[][Raw]{\includegraphics[width=.2\columnwidth]{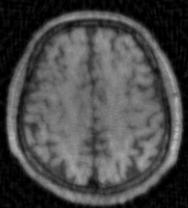}}\hfill%
    \subfloat[][FISTA]{\includegraphics[width=.2\columnwidth]{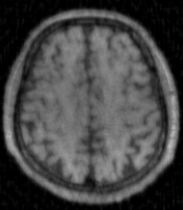}}\hfill%
    \subfloat[][FISTA-Net]{\includegraphics[width=.2\columnwidth]{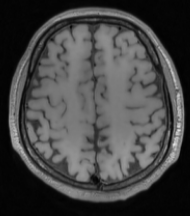}}
    \caption{Comparison of classical FISTA and model-based FISTA-Net reconstruction of brain MR images. These images were taken from \cite{aromal2024fista}.}
    \label{fig:fistanet}
\end{figure}

Naturally, there is a long line of research into proper ways of reconstructing medical imaging data~\cite{roemer1990nmr,kawata2007constrained,block2007undersampled,beck2009fast}.
These ``classical'' algorithms are based on mathematical optimization problems and have attractive theoretical guarantees,
such as convergence rates and bounds on the reconstruction error.
More recently, the use of neural networks to reconstruct medical imaging data has been explored~\cite{aggarwal2018modl,hammernik2018learning,sriram2020end,aromal2024fista}.
Several studies have confirmed that the resulting reconstructions are more visually appealing than those generated by classical algorithms~\cite{muckley2020state,muckley2021results}.
As an example, \cref{fig:fistanet} compares the classical FISTA algorithm~\cite{beck2009fast} to the more recent model-based FISTA-Net method~\cite{aromal2024fista}. The improvement in visual quality is obvious here.

\begin{figure}
    \subfloat[][]{
        \includegraphics[width=.2\columnwidth]{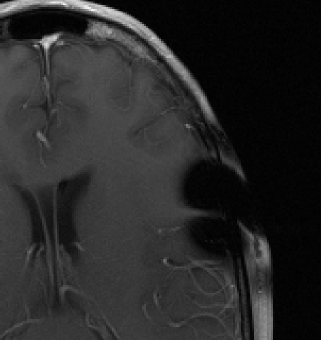}\hspace{.2cm}
        \includegraphics[width=.2\columnwidth]{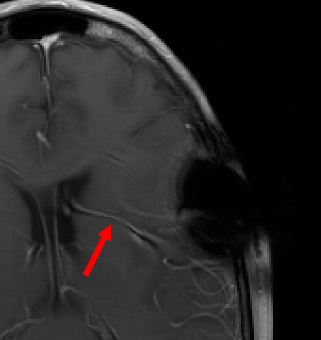}
    }
    \hfill
    \subfloat[][]{
        \includegraphics[width=.2\columnwidth]{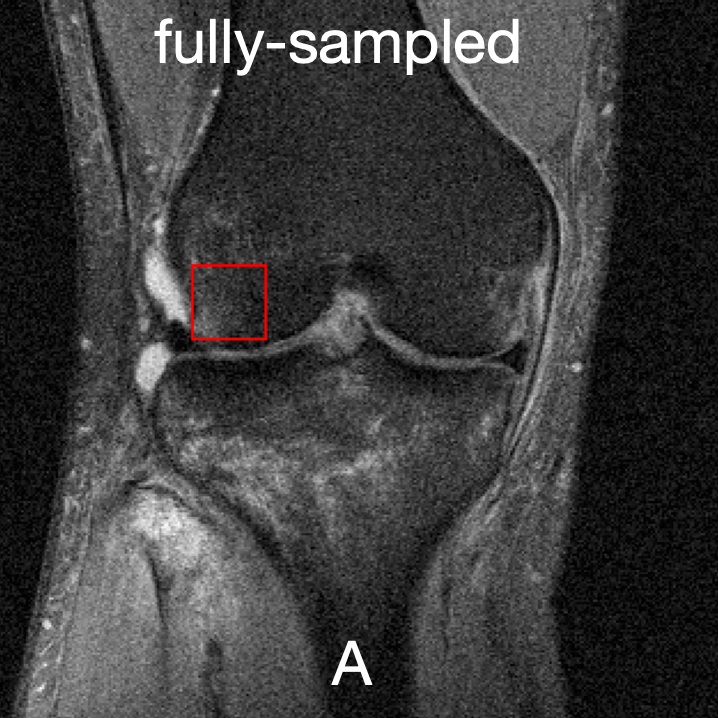}\hspace{.2cm}
        \includegraphics[width=.2\columnwidth]{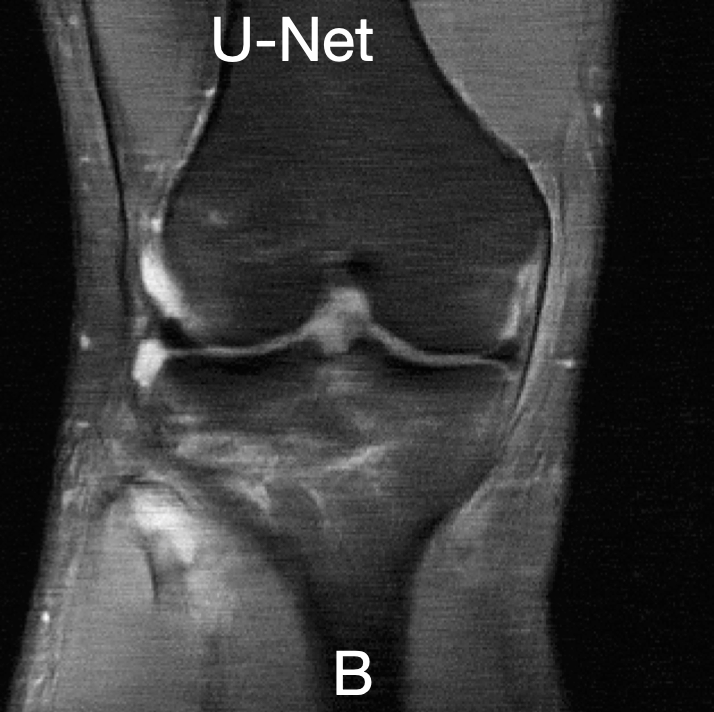}
    }
    \caption{Examples of hallucinations introduced by model-based reconstructions of MR images.
    (a) In this image from \cite{muckley2020state}, the model-based reconstruction introduces an additional sulcus in the brain which was not present in the original.
    (b) In this image taken from \cite{cheng2020addressing}, the model-based reconstruction removes evidence of a meniscal tear that is present in the original data.}
    \label{fig:hallucinations}
\end{figure}

Nevertheless, generative neural networks typically lack theoretical guarantees, calling their reliability in this context into question~\cite{sidky2020cnns,gottschling2025troublesome}.
Empirically, it is known that generative models of any kind are susceptible to so-called \emph{hallucinations}.
Hallucinations occur when a generative model adds an important feature to its output that was not supposed to be there, or removes a feature that should have been present.
This problem is particularly well-known in the field of language modeling: modern large language models (LLMs) are notoriously prone to generating convincing misinformation or outright nonsense~\cite{farquhar2024detecting,xu2024hallucination}.
However, imaging models are not exempt from this problem: a few examples are shown in \cref{fig:hallucinations}. These alterations can have a significant impact on patient care. As such, it is an important open problem to detect when such hallucinations may have occurred and to prevent them from happening if possible. Currently, there are no effective methods to address these issues~\cite{heckel2024deep}.

In this work, we aim to contribute towards effective mitigation of hallucinations in medical imaging by crafting an algorithm which actively \emph{provokes} them in state-of-the-art model-based reconstructions.
Our motivation is that such an algorithm would facilitate follow-up research into the detection and mitigation of hallucinations.
For instance, the algorithm could be used to create a data set of reconstructions containing hallucinations. Combined with clean reconstructions, a training data set can be created to train a detector that flags hallucinations. Alternatively, the algorithm could also serve as a component in an adversarial training regime that improves the robustness of generative models such that hallucinations occur less frequently.

We evaluate our attack on the fastMRI data set~\cite{zbontar2018fastmri} against UNet~\cite{ronneberger2015unet} and E2E-VarNet~\cite{sriram2020end} models, which have been found to obtain good performance on these reconstruction tasks.
Our attack seems to be highly successful at eliciting hallucinations via imperceptibly small perturbations of the input, leading to the conclusion that these models are highly unstable and hallucinations can easily arise in practice purely as a result of noise.
We perform further experiments to attempt to detect the hallucinations using standard image quality metrics, such as peak signal-to-noise ratio (PSNR), normalized root mean squared error (NRMSE) and structural similarity index (SSIM; \cite{wang2004image}).
We find that the distributions of these metrics overlap almost completely and hence cannot be used to distinguish reliable reconstructions from those containing hallucinations.
The code to reproduce our experiments is available at \codeurl.

\section{Background}
Many imaging reconstruction problems are essentially instances of a particular \emph{linear inverse problem}~\cite{tropp2010computational,ribes2008linear}. Formally, we observe some noisy vector $\bm{y} \in \comp^m$ which is assumed to be produced according to a so-called \emph{forward model}:
\begin{equation}\label{eq:linv}
    \bm{y} = \bm{\Phi} \cdot \bm{x}^\natural + \bm{e}.
\end{equation}
Here, $\bm{x}^\natural \in \comp^n$ is the true signal, $\bm{\Phi} \in \comp^{m \times n}$ is a measurement matrix (also referred to as the \emph{forward transform}) and $\bm{e} \in \comp^m$ is noise. We always have $m < n$, so the original signal $\bm{x}^\natural$ is subject to both compression (via $\bm{\Phi}$) and additive noise (given by $\bm{e}$). The goal is to reconstruct $\bm{x}^\natural$ as faithfully as possible given the noisy and compressed observations $\bm{y}$.

MRI and CT scans present typical examples of linear inverse problems. In CT, the observations $\bm{y}$ are sinograms obtained via X-ray projections of the original image $\bm{x}^\natural$ from various directions~\cite{zhang2022use}. In the case of MRI, the observations are spatial frequencies in Fourier space, referred to as \emph{k-space} in the literature~\cite{heckel2024deep}. Mathematically, the measurement matrices of CT and MRI are given by the discrete Radon transform~\cite{beylkin1987discrete} and discrete Fourier transform~\cite{winograd1978computing}, respectively.

The solutions to linear inverse problems are typically obtained via iterative optimization procedures based on a probabilistic model of the noise and the forward transform. The conditional density $p_{\bm{Y} \mid \bm{X}}(\bm{y} \mid \bm{x})$ is usually known from the physics of the measurement process. The most common choice is to assume Gaussian noise, so that $\bm{y}$ is normally distributed around $\bm{\Phi} \cdot \bm{x}^\natural$. This leads to an optimization problem known as \emph{basis pursuit}~\cite{chen1994basis}:
\begin{equation}\label{eq:inv}
    \hat{\bm{x}} = \arg\min_{\bm{x} \in \comp^n}~\frac 12\norm{\bm{y} - \bm{\Phi} \cdot \bm{x}}_2^2 + \lambda\reg(\bm{x}).
\end{equation}
The $L_2$ penalty is referred to as the \emph{data consistency objective} and is a direct result of the Gaussian assumption on $\bm{e}$. The second term, $\reg(\bm{x})$, is a regularizer which depends on the choice of prior $p_{\bm{X}}(\bm{x})$. Common choices for $\reg$ include the $L_1$ norm in image space, $\norm{\bm{x}}_1$; the $L_1$ norm in wavelet space, $\norm{\bm{\Psi} \cdot \bm{x}}_1$, where $\bm{\Psi}$ is a discrete wavelet transform~\cite{torrence1998practical}; or the total variation of $\bm{x}$~\cite{chambolle2010introduction}. The resulting reconstruction algorithms are well-documented in various surveys and textbooks~\cite{foucart2013}.

\subsection{Model-based reconstruction}
The success of deep learning approaches in the image domain suggests that deep neural networks may have potential in determining good solutions to linear inverse problems for medical imaging.
While it is possible to train a neural network to directly learn the mapping from the observation space to the original image space~\cite{hammernik2018learning}, this strategy poses the risk of hallucinations~\cite{heckel2024deep,gottschling2025troublesome,bhadra2021hallucinations} and lacks formal guarantees on the reconstruction quality.
In particular, when a neural network is used to reconstruct the image directly, it is not clear what the underlying inverse problem actually is that is being solved, which casts doubt on the reliability of the results~\cite{sidky2020cnns}.

In light of these issues, some authors have attempted to use deep networks as a regularizer rather than a full reconstruction algorithm.
Specifically, they retain the original optimization problem \eqref{eq:inv} but incorporate the neural network into the regularizer $\reg(\bm{x})$.
This is the approach taken by the MoDL framework~\cite{aggarwal2018modl}, which reconstructs the image by solving \eqref{eq:inv} with
\[
    \reg(\bm{x}) = \norm{\bm{x} - F_\theta(\bm{x})}_2^2,
\]
where $F_\theta$ is a trained denoising neural network.
In this case, the regularization consists of minimizing the estimated noise in the image as determined by a neural network reconstruction.

\subsection{Adversarial perturbations}
Adversarial perturbations are worst-case alterations of the inputs to a machine learning model, in the sense that they are intended to be imperceptible or harmless to a human observer while causing models to make mistakes~\cite{szegedy2013intriguing}. In the image domain, they typically take the form of (nearly) imperceptible distortions. For discriminative models, the goal of adversarial perturbations is usually to cause misclassification. In the case of generative models, performance is generally harder to quantify but usually relies on minimizing some measure of similarity between the output of the model and a reference. Examples include the mean squared error (MSE) or the structural similarity index measure (SSIM)~\cite{wang2004image}.

Adversarial perturbations have been the subject of much research for a long time now~\cite{biggio2018wild,biggio2013evasion}, and while certain promising defenses have been proposed, the problem is still far from being considered ``solved'' in any meaningful way~\cite{peck2023introduction,costa2024deep}. In this work, we will make use of adversarial perturbations to craft imperceptible distortions that, when added to an MR image, result in hallucinations when reconstructed by a neural network.

\subsection{Related work}
Adversarial perturbations for medical images are not new, and many attacks and defenses have been proposed to date~\cite{dong2024survey,kaviani2022adversarial}. However, the vast majority of existing work focuses on classification tasks, which is the most straightforward setting for an adversarial attack because success is easy to measure through misclassification rate. By contrast, we focus here on attacking \textit{reconstruction} tasks, which is a generative setting where the objective is to remove undesirable artifacts from the data and improve visual quality.

In this vein, \cite{darestani2021measuring} also consider the UNet and E2E-VarNet models we study here and compare their robustness to classical methods such as total variation (TV) minimization.
They find that \emph{both} neural networks as well as classical reconstruction algorithms are vulnerable to adversarial perturbations specifically tailored to the methods in question.
This finding seems to contradict other studies such as \cite{genzel2022solving} who claim that TV minimization is provably robust.
In fact, \cite{genzel2022solving} further find that neural networks are resilient against adversarial perturbations, which also contradicts our own results.
Although the existing body of work on robustness of image reconstruction seems to favor the view that classical reconstruction algorithms are stable whereas neural networks are not, consensus on this question does not seem to have been reached yet.

Theoretical results on the stability of classical reconstruction algorithms have been obtained by \cite{del2023stability}.
From their work, we can conclude that classical methods should not suffer from the same instability with respect to small perturbations, given the favorable continuity properties these methods enjoy.
\cite{colbrook2022difficulty} prove several results related to the (non-)existence of accurate and stable neural networks for inverse problems.
They also introduce FIRENETs, a family of neural networks that can be efficiently trained and have provable stability properties.
More recently, \cite{gottschling2025troublesome} related the instabilities of model-based reconstruction methods to the kernel of the measurement matrix.
Their main results are broadly applicable to many reconstruction algorithms (not merely neural networks), and consist of a few ``no free lunch'' theorems which show that (paradoxically) higher performance increases the probability of hallucination, and that hallucinations are not rare events (in the sense of having non-zero probability of occurring).

The work that comes closest to ours is \cite{morshuis2022adversarial}, who also study adversarial perturbations for MRI reconstruction.
They evaluate adversarial attacks against UNet~\cite{ronneberger2015unet} and E2E-VarNet~\cite{sriram2020end} on the fastMRI data set~\cite{zbontar2018fastmri}.
These attacks are untargeted in the sense that they aim to maximize the reconstruction error within a specified region with respect to fully sampled reference images.
By contrast, in this work, we study a targeted adversarial attack which aims to insert a specific detail into the reconstruction that is not actually present.
This attack requires no reference data and supports any type of detail which can be efficiently rendered onto an image.
We also perform additional experiments in an attempt to detect the hallucinations we inserted using standard image quality metrics, such as PSNR, NRMSE and SSIM, and find that these are ineffective.

\section{Method}
The goal of our attack is to craft invisible perturbations such that model-based reconstructions of the perturbed data points exhibit distortions that could mislead diagnostic interpretation. To achieve this, given a $k$-space data vector $\bm z \in \comp^n$ and a reconstruction map $F: \comp^n \to \real^n$, we solve the following optimization problem:
\begin{equation}\label{eq:attack}
    \begin{aligned}
        \bm\delta^\star = \arg\min_{\bm\delta \in \comp^n}~&\frac{\|\bm m \odot (F(\bm z + \bm\delta) - \bm y_t)\|_2^2}{\|\bm m\|_1}\\
        &+ \frac{\|(1 - \bm m) \odot (F(\bm z + \bm\delta) - F(\bm z))\|_2^2}{\|1 - \bm m\|_1}
    \end{aligned}
\end{equation}
subject to the constraint $\norm{\bm\delta}_\infty \leq \varepsilon$, where $\varepsilon > 0$ is the perturbation budget,
$\bm m \in \real^n$ is a binary mask, and $\bm y_t$ is a target reconstruction.

The budget $\varepsilon$ must be sufficiently small so that the generated perturbations are invisible to the naked eye,
but large enough to allow insertion of the desired hallucinations in the reconstruction.
The specific value therefore depends on the application; see \cref{sec:experiments} for further details.

The target reconstruction $\bm y_t$ is designed by drawing a short white line in the center of the original reconstruction $F(\bm z)$.
The mask $\bm m$ delimits the region where this line was added.
In this way, the objective function in \eqref{eq:attack} strikes a balance between the insertion of the specified detail on the one hand and faithful reconstruction of the image on the other:
the first term in the loss penalizes the difference between the reconstruction and the target image,
whereas the second term penalizes the difference between the reconstruction and the ground truth.
The optimal solution is achieved when the perturbation $\bm\delta$ is such that $F(\bm z + \bm\delta)$ is identical to $\bm y_t$ within the target region defined by the mask $\bm m$ and identical to $F(\bm z)$ outside the target region.

Pseudocode describing the attack is provided in \cref{app:pseudocode}.
The method is inspired by the basic iterative method (BIM), which is a popular algorithm for generating adversarial examples~\cite{kurakin2018adversarial}.
Note that $k$-space data is complex-valued, but we consider only real-valued perturbations $\bm\delta$ in our algorithm.
This is done to simplify the computations on the one hand, but also because we noticed that, for our purposes, it suffices to only perturb the real components of the $k$-space vector.
Hence we leave the imaginary components of all samples unchanged.

Although in this work we only consider lines drawn onto the center of the image, the attack we developed here is agnostic to the specific target.
Indeed, the attack can use any target $\bm y_t$ with associated mask $\bm m$
and hence can be used to provoke all kinds of hallucinations, as long as they can be efficiently rendered onto an image.
We chose lines here because they are simple yet already highly effective at introducing unwanted artifacts,
but other shapes such as squares or circles may also be used.
An interesting further experiment could be to use parts of other images, which include real pathologies,
and render them onto images which do not contain them.
In this way, a convincing meniscal tear might be hallucinated onto an image of a healthy knee, for instance.
This can be facilitated using annotations of pathologies in existing MRI data sets, such as fastMRI+.\footnote{\url{https://github.com/microsoft/fastmri-plus}}

\section{Experiments}\label{sec:experiments}
We evaluated our attack algorithm on the fastMRI data set~\cite{zbontar2018fastmri} using a pretrained UNet~\cite{ronneberger2015unet} and E2E-VarNet~\cite{sriram2020end}.
We consider single-coil and multi-coil knee images, as well as multi-coil brain images.
We do not evaluate E2E-VarNet on the single-coil knee images, since there were no pretrained checkpoints available for that model on this part of the fastMRI data set.
The adversarial perturbations are generated by solving \eqref{eq:attack} using \cref{alg:attack} with $T = 150$ iterations.
The magnitude of the perturbation was bounded in the $L_\infty$ norm to a maximum of $\varepsilon = \num{1e-6}$ with a step size of $\alpha = \num{1e-7}$.

\subsection{Results}
We performed a quantitative evaluation of the results using peak signal-to-noise ratio (PSNR), normalized root mean squared error (NRMSE), and structural similarity index measure (SSIM).
These metrics are computed separately for the pairs of original and perturbed input samples, as well as the pair of original and perturbed reconstructions.
We also investigated the value of the objective function in \eqref{eq:attack} and found that it was almost always negligibly small across all models and data sets (on the order of $10^{-8}$ or less). We therefore do not include these values in the report, but we conclude that they are indicative of a successful attack.

\begin{table}[ht]
    \caption{Summary statistics of the experiments. The reported numbers are means over the entire data set along with their standard deviations in parentheses.}
    \centering
    \begin{tabular}{p{1.5cm}p{2cm}p{2cm}p{2cm}p{2cm}}
        \toprule
        Model & Data & PSNR $\uparrow$ & NRMSE $\downarrow$ & SSIM $\uparrow$\\
        \midrule
        \multirow[t]{3}{*}{UNet}
            & sc-knee
                & 55.60 (6.18)  & 0.01 (0.01) & 1.00 (0.00)\\
            &   & 34.24 (10.10) & 0.17 (0.33) & 0.95 (0.10)\\
            \cmidrule(l){2-5}
            & mc-knee
                & 50.35 (5.74) & 0.02 (0.01) & 1.00 (0.00)\\
            &   & 28.48 (4.62) & 0.23 (0.13) & 0.82 (0.13)\\
            \cmidrule(l){2-5}
            & mc-brain
                & 61.72 (4.58)  & 0.01 (0.00) & 1.00 (0.00)\\
            &   & 26.00 (11.06) & 0.76 (1.07) & 0.68 (0.34)\\
        \midrule
        \multirow[t]{2}{*}{VarNet}
            & mc-knee
                & 50.69 (6.24) & 0.02 (0.01) & 1.00 (0.01)\\
            &   & 31.70 (1.70) & 0.42 (0.18) & 0.93 (0.03)\\
            \cmidrule(l){2-5}
            & mc-brain
                & 60.10 (4.44) & 0.01 (0.00) & 1.00 (0.00)\\
            &   & 41.72 (7.48) & 0.10 (0.10) & 0.98 (0.03)\\
        \bottomrule
    \end{tabular}
    \label{tab:results}
\end{table}

Summary statistics of these metrics are given in \cref{tab:results}, with values rounded to two decimal places.
For every metric $M \in \{ \mathrm{PSNR}, \mathrm{NRMSE}, \mathrm{SSIM} \}$, we report the mean and standard deviation of two values across the data set:
\begin{enumerate}
    \item The metric computed on the zero-filled\footnote{Zero-filling is the most basic reconstruction method where we simply apply an inverse Fourier transform on the raw $k$-space data (possibly padded with additional zeros), without making any attempt at reducing noise.} original and adversarial input samples, $M(\ZF(\bm z), \ZF(\bm z + \bm\delta))$, where $\ZF(\cdot)$ denotes the zero-filling operation.
    
    \item The metric computed on the reconstructions, $M(F(\bm z), F(\bm z + \bm\delta))$.
\end{enumerate}
We report these two values for each metric to gauge the success of the attack: in each case, we expect the metric computed on the zero-filled pair to be ``good'' -- high in the case of PSNR and SSIM, low for NRMSE -- but the reconstructions should be noticeably worse.
We do indeed observe this pattern in \cref{tab:results}: the metric values are consistently better for the zero-filled originals compared to the reconstructions.
For the originals, PSNR values are always at least 50 dB, the NRMSE is 0.02 or less and the SSIM is almost 100\%.
This suggests that the clean input samples and the adversarial samples are visually indistinguishable, as desired.
On the other hand, the metric values for the reconstructed samples are noticeably worse:
PSNR values drop to around 40 dB or less, NRMSE increases by at least an order of magnitude and SSIM values can drop up to thirty percentage points.
Although these values still indicate that the reconstructions are similar, there is a significant difference.
Moreover, it is expected for the reconstructions to retain much similarity, because the artificial detail we added using our adversarial attack
only covers a small portion of the image.

\begin{figure}[ht]
    \begin{center}
        \subfloat[][sc-knee]{
            \includegraphics[width=.4\columnwidth]{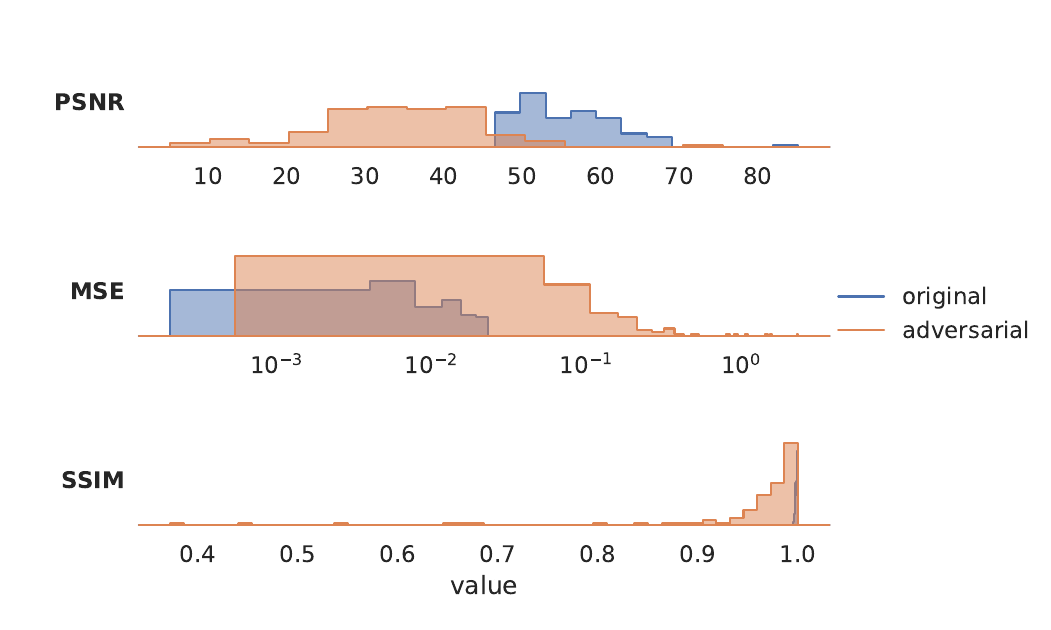}
        }
        \subfloat[][mc-knee]{
            \includegraphics[width=.4\columnwidth]{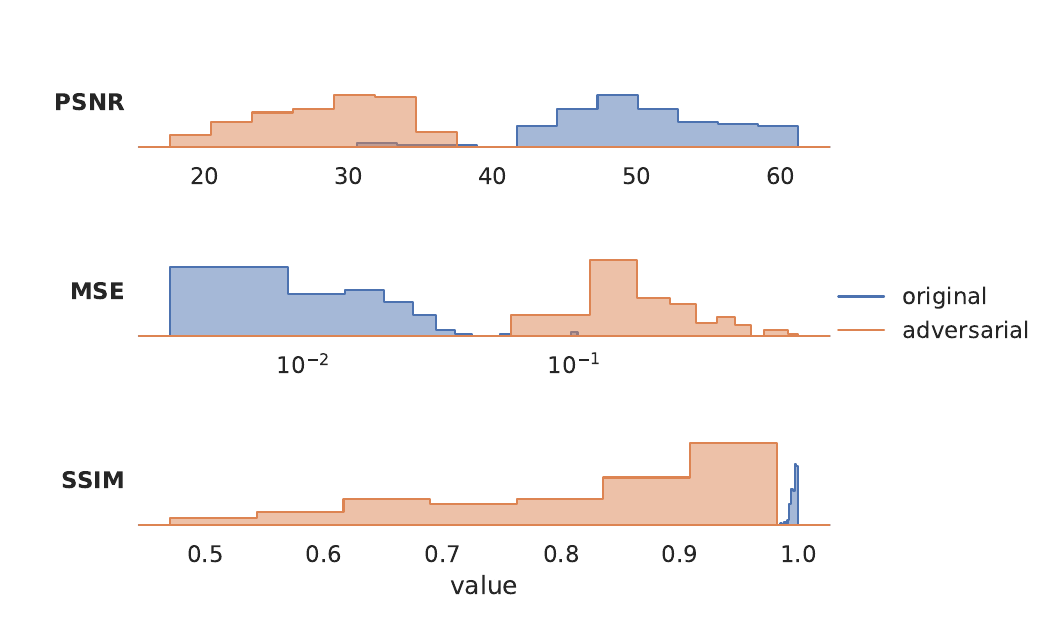}
        }
        
        \subfloat[][mc-brain]{
            \includegraphics[width=.4\columnwidth]{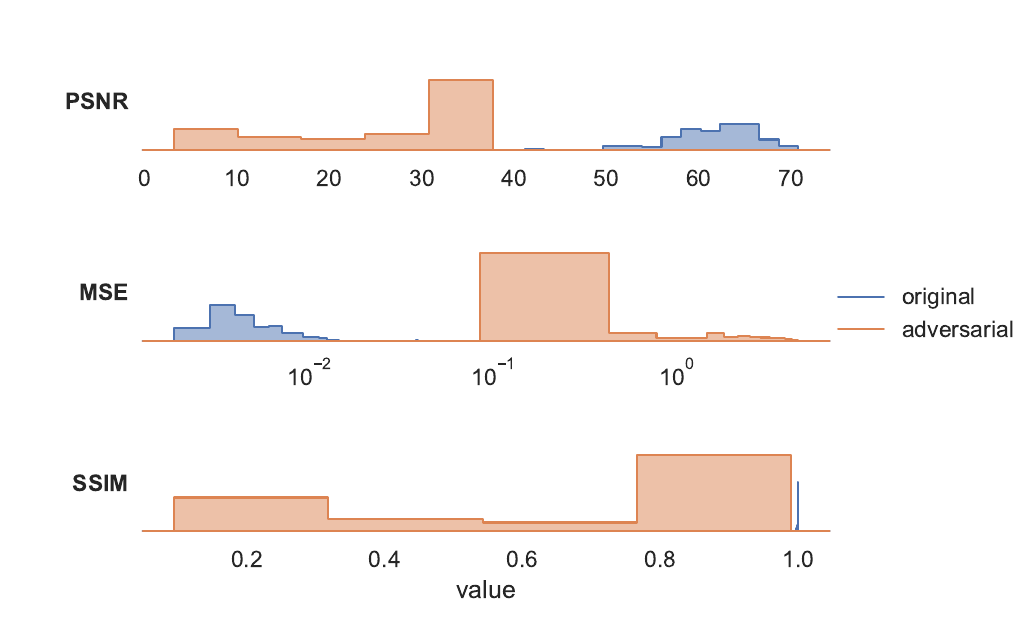}
        }
    \end{center}
    \caption{Distributions of the metric values for the UNet model.}
    \label{fig:unet}
\end{figure}

\begin{figure}[ht]
    \begin{center}
        \subfloat[][mc-knee]{
            \includegraphics[width=.4\columnwidth]{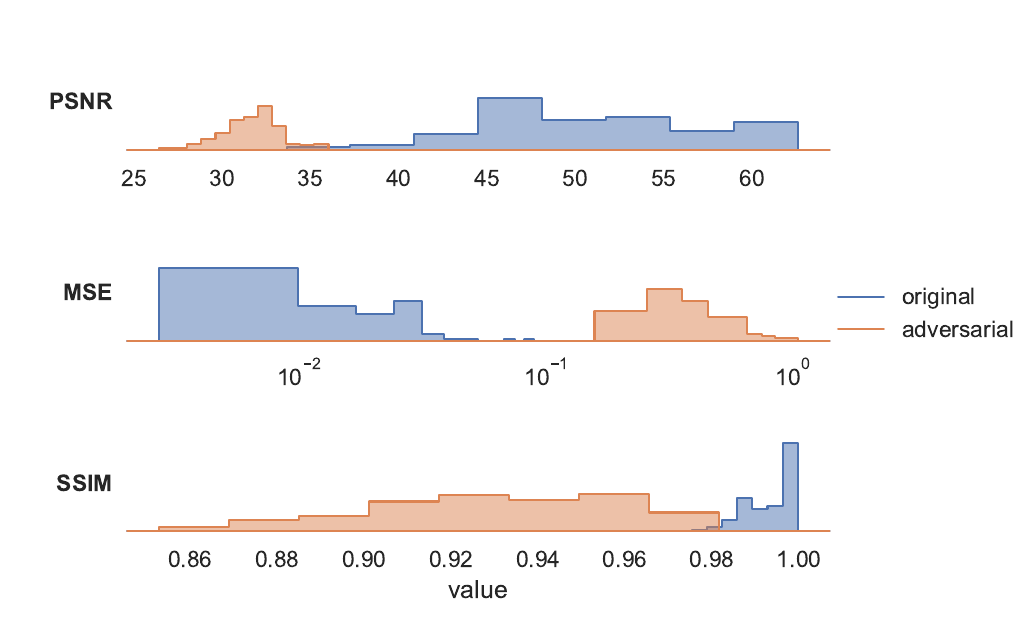}
        }
        \subfloat[][mc-brain]{
            \includegraphics[width=.4\columnwidth]{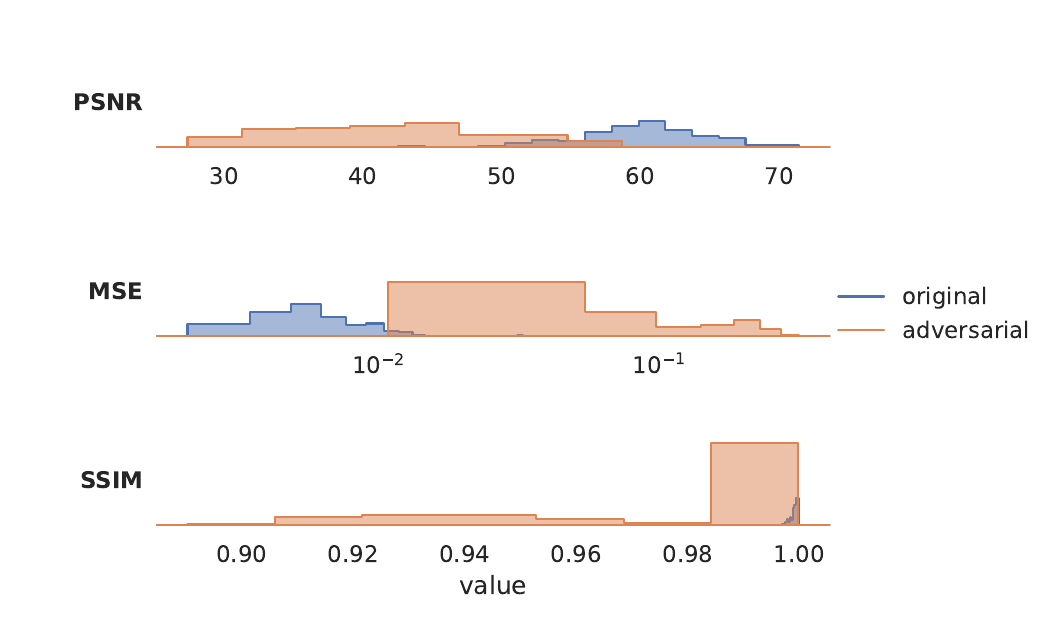}
        }
    \end{center}
    \caption{Distributions of the metric values for the E2E-VarNet model.}
    \label{fig:varnet}
\end{figure}

To further investigate the extent to which our attack can be called successful,
we plot the full distributions of the metric values in \cref{fig:unet} for the UNet model and \cref{fig:varnet} for the VarNet model.
These values are visualized as histogram-based ridge plots, showing the overlap between the distributions.
Note that the MSE values are shown on a semi-logarithmic scale.

Consistent with \cref{tab:results}, we observe that the distributions of all metric values tend to be well-separated between the original samples and the reconstructions.
Some exceptions to this can be observed: the distributions of NRMSE for the sc-knee data reconstructed by the UNet model largely overlap,
as do the distributions of SSIM values for the sc-knee data reconstructed by the UNet model and mc-brain data reconstructed by the E2E-VarNet model.
In all of these cases, however, we observe that the metrics involved have a significantly larger spread after reconstruction.
This indicates that the hallucinated details may not always be very visible, but given the large variance in the values, the attack
is still very likely to produce noticeable distortions for the majority of inputs.

A qualitative assessment of the generated images was also performed but has been deferred to \cref{app:qualitative} due to space constraints.
From those results, we conclude that the adversarially perturbed samples can lead to realistic reconstructions which exhibit biologically plausible distortions that could mislead expert interpretation, and that the insertion of the artificial detail will often cause further distortion far beyond the target region delineated by our mask.

\subsection{Detecting hallucinations}
In this section, we turn to the question of detectability of our generated hallucinations.
We consider a simple threat model where an adversary may have contaminated part of a data set using perturbations generated by \cref{alg:attack}.
The defender is given access only to $k$-space vectors $\bm z_1, \ldots, \bm z_N$ and must decide which (if any) of the samples have been corrupted.

As argued above, it is expected that classical reconstruction algorithms such as total variation, conjugate gradients and FISTA do not suffer from hallucinations.
It therefore stands to reason that such algorithms may be used to compute an expected distribution of residual reconstruction errors
against which we may test the model-based reconstructions.
Specifically, given a potentially perturbed $k$-space vector $\bm z$, we ask whether it is possible to detect hallucinations using standard image quality metrics
applied to the pair $R(\bm z)$ and $F(\bm z)$, where $R(\cdot)$ denotes some classical reconstruction algorithm.

We test this hypothesis by first using \cref{alg:attack} to generate a data set of perturbed $k$-space vectors $\tilde{\bm z}_1, \ldots, \tilde{\bm z}_N$.
Then, for each unperturbed $k$-space vector $\bm z$ and its adversarially perturbed counterpart $\tilde{\bm z}$,
we compute PSNR, NRMSE and SSIM metrics on the pairs of reconstructions $(\tv(\bm z), F(\bm z))$ and $(\tv(\tilde{\bm z}), F(\tilde{\bm z}))$, where $\tv(\cdot)$ is the total variation reconstruction algorithm as implemented in the SigPy library\footnote{\url{https://github.com/mikgroup/sigpy}} using sensitivity maps estimated by ESPIRiT~\cite{uecker2014espirit}.
If these distributions are well-separated, then the defender can simply compute these metrics for each sample in the data set and compare them to a pre-determined threshold.

\begin{figure}[ht]
    \begin{center}
        \subfloat[][sc-knee]{
            \includegraphics[width=.4\columnwidth]{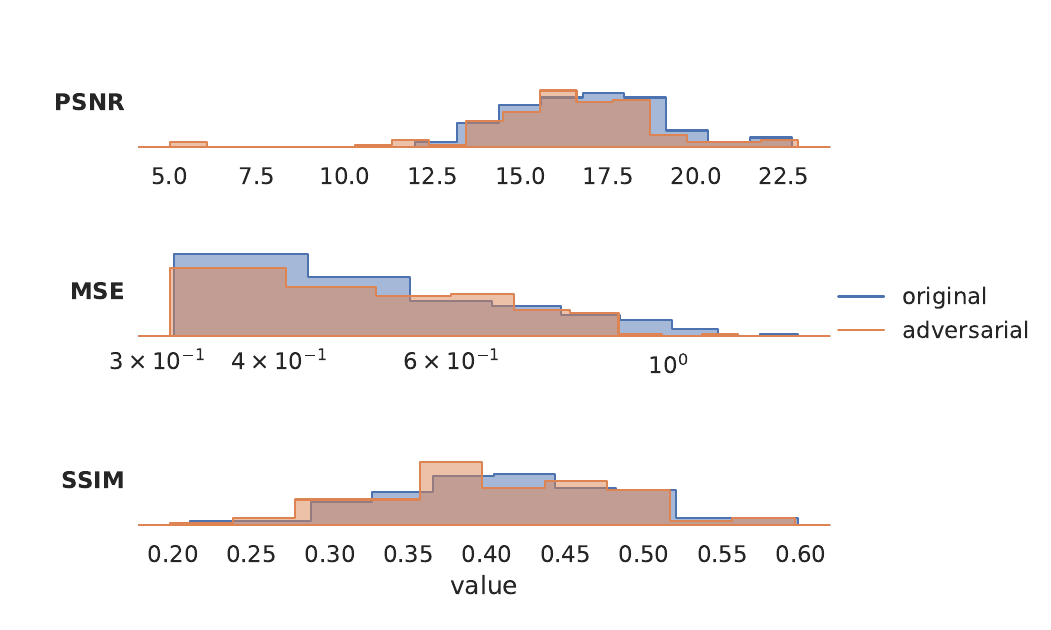}
        }
        \subfloat[][mc-knee]{
            \includegraphics[width=.4\columnwidth]{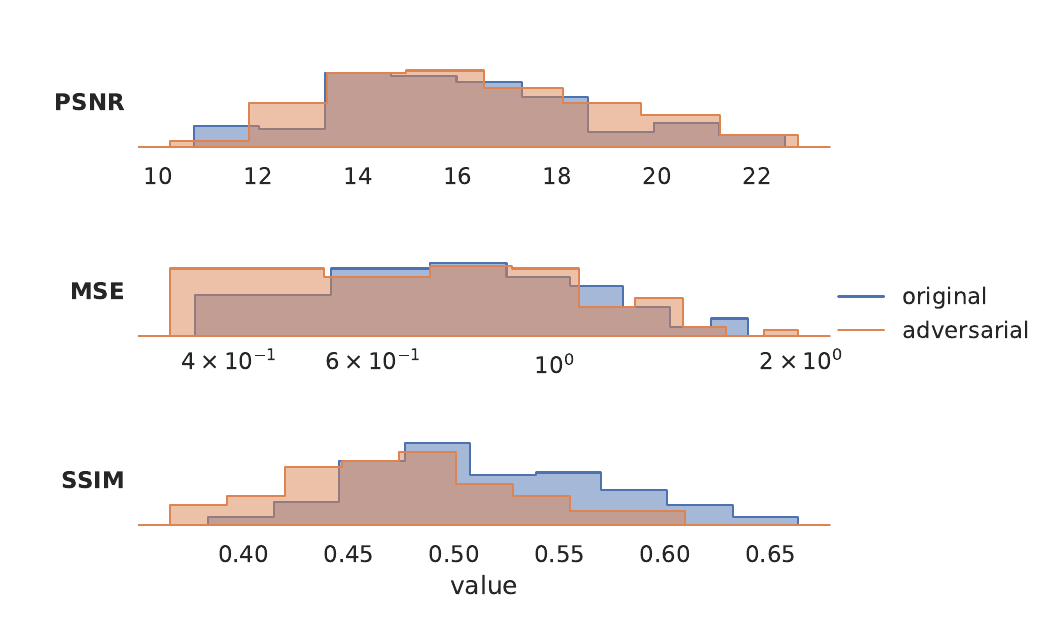}
        }
    
        \subfloat[][mc-brain]{
            \includegraphics[width=.4\columnwidth]{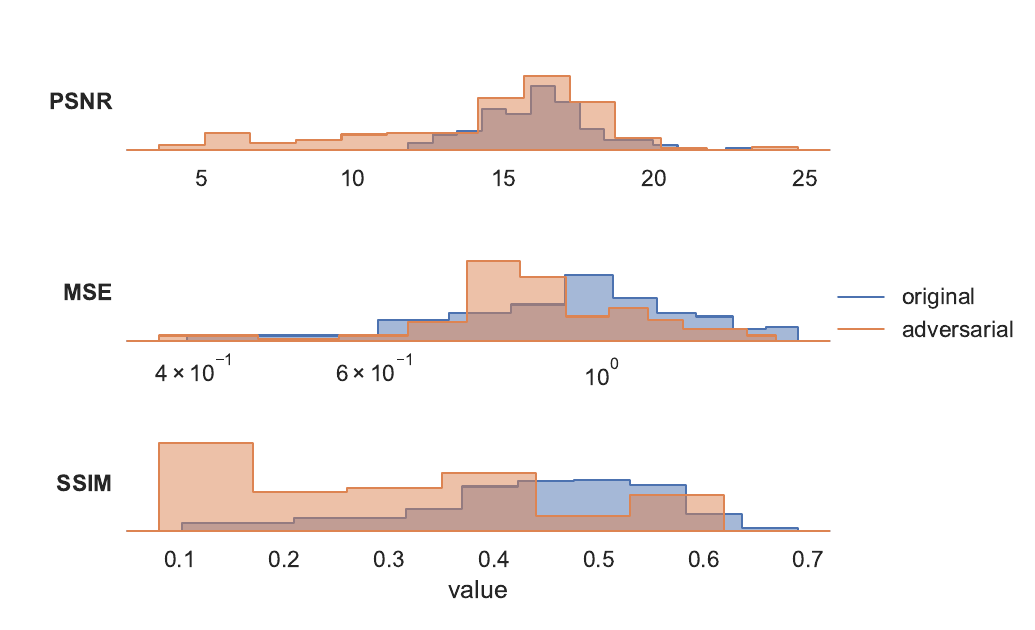}
        }
    \end{center}
    \caption{Distributions of the metric values for the UNet model using total variation reconstructions.}
    \label{fig:unet-tv}
\end{figure}

\begin{figure}[ht]
    \begin{center}
        \subfloat[][mc-knee]{
            \includegraphics[width=.4\columnwidth]{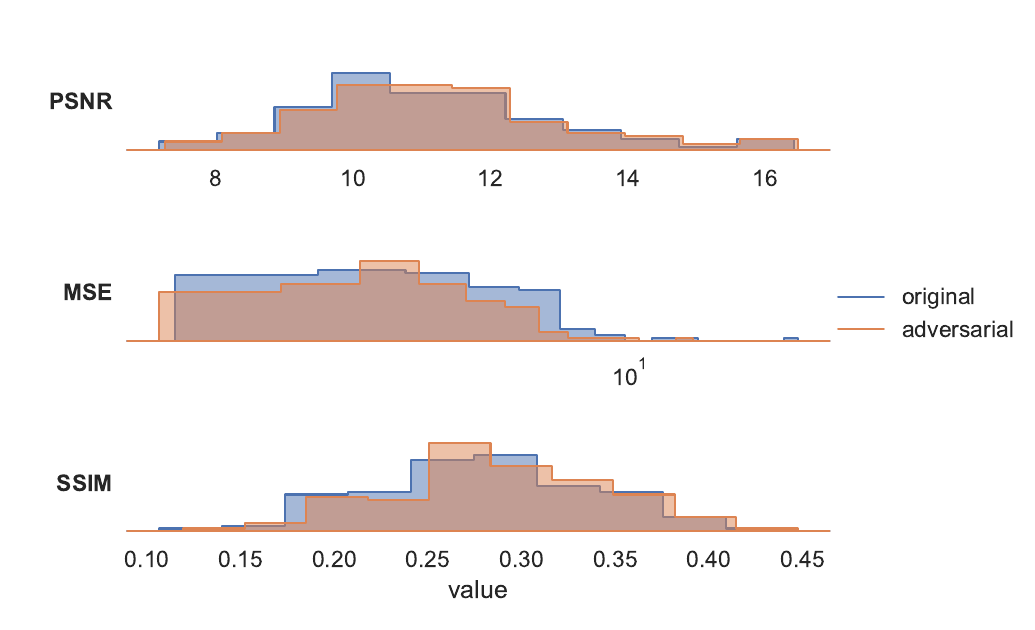}
        }
        \subfloat[][mc-brain]{
            \includegraphics[width=.4\columnwidth]{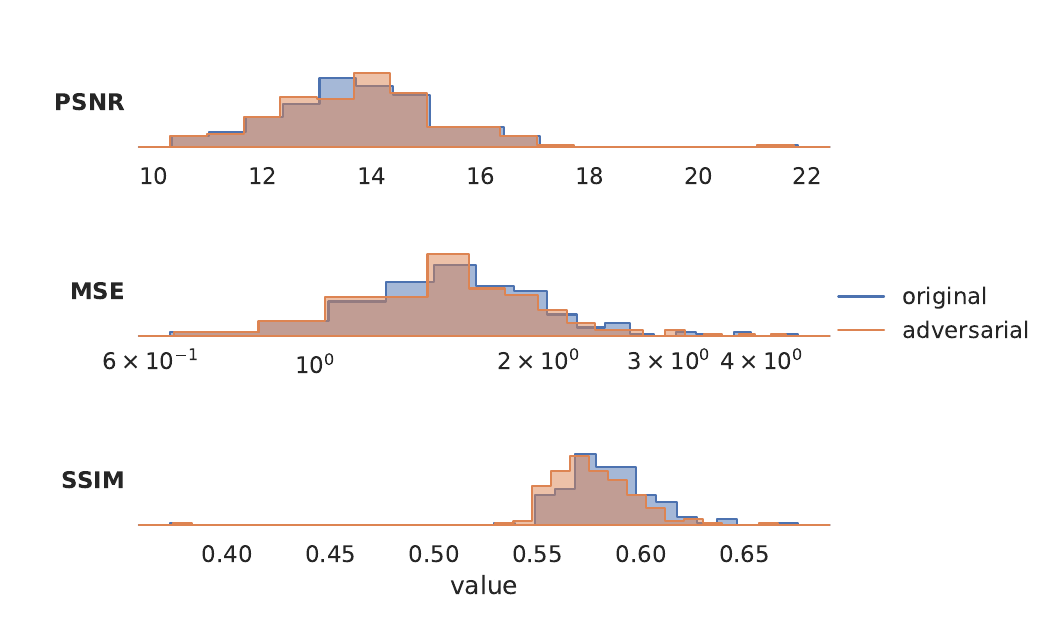}
        }
    \end{center}
    \caption{Distributions of the metric values for the E2E-VarNet model using total variation reconstructions.}
    \label{fig:varnet-tv}
\end{figure}

The distributions of the resulting values are given in \cref{fig:unet-tv} for the UNet model and \cref{fig:varnet-tv} for the E2E-VarNet model.
From these results we observe that, for every metric, the two distributions almost entirely overlap.
The only exception seems to be the SSIM metric on the mc-brain data for the UNet model, which exhibits an unusual peak near 10\%.
However, this pattern never repeats for any of the other data sets and does not appear at all for the E2E-VarNet model.
It therefore seems unreliable, and the distributions still overlap sufficiently to make any attempt at separation using the SSIM prone to error.

We stress that \cref{alg:attack} was not explicitly constructed to maximize overlap between these metrics.
Indeed, our attack merely minimizes \eqref{eq:attack} and does not make any use of other reconstruction algorithms, nor does it directly optimize PSNR or SSIM.
Yet it appears this approach already suffices to make the generated hallucinations essentially undetectable using standard image quality metrics.
As is common practice in adversarial attacks~\cite{tramer2020adaptive}, \cref{alg:attack} may be extended by including
such additional objectives into the loss function in order to make the perturbations even less detectable, but it seems this is not necessary here.

Note that we do not explore the possibility of training model-based detectors for these perturbations here.
This omission is due to our expectation that such approaches are bound to be ineffective, since
the field of adversarial machine learning has learned through experience~\cite{peck2023introduction,tramer2020adaptive,athalye2018obfuscated,carlini2017adversarial}
that adversarial attacks can be easily adapted to incorporate the outputs of such detectors.
Hence, an evaluation where we train a neural network to differentiate between clean and perturbed $k$-space vectors would
merely serve to create a false sense of security, and would likely result in a swift break of the detector in a follow-up work.

These findings suggest that, in order to solve the problem of hallucinations in model-based MRI reconstructions,
mathematically principled approaches will be needed that give rise to \emph{certified} defenses or detectors.
In this vein, carefully constructed adversarial training regimes~\cite{lee2021towards,mao2023connecting} or the application of denoised smoothing~\cite{cohen2019certified,salman2020denoised} may be explored.

\section{Conclusions}
We have proposed an adversarial attack which can insert hallucinations into model-based MRI reconstructions via invisible perturbations of the input samples.
We evaluated the algorithm using several image quality metrics and found that it often succeeds in causing significant distortion of the reconstructed samples while remaining invisible to the naked eye in input space.
Our results imply that, in the absence of reference data, such distortions are not easily detected using ``traditional'' metrics such as peak signal-to-noise ratio, mean squared error or structural similarity index.
Moreover, a qualitative assessment of the results indicates that whenever our attack is successful, it tends to cause further biologically plausible distortions beyond the neighborhood of the artificially inserted detail.

We conclude that state-of-the-art model-based MRI reconstruction algorithms are highly unstable with respect to small noise and can easily be induced to insert hallucinatory structures.
This poses a risk when these models are used in medical contexts, where such perturbations may arise spontaneously through noise and may affect diagnosis and eventual treatment of patients.
Future work in this area should focus on reducing the instability of reconstruction maps to small perturbations as well as developing detection mechanisms which can signal when reconstructions may be unreliable.
The attack introduced in this work may form the basis for an adversarial training regime that could improve the robustness of model-based MRI reconstruction
or serve as a baseline against which to benchmark potential detection approaches.

Given the historical failure of many approaches to detecting adversarial perturbations, we argue that future research in this direction
should focus on mathematically principled techniques that lead to detectors with provable correctness even in the presence of worst-case perturbations.
It is known that detecting adversarial perturbations is a very hard problem in general,
and no satisfying general-purpose solutions exist at this time~\cite{peck2023introduction,biggio2018wild}.
However, when the problem is constrained specifically to the detection of hallucinations in MRI reconstruction,
existing mathematical results in compressed sensing may be employed which are not available in more general settings~\cite{peck2024robust,chen2013robust,sulam2020adversarial}.
It is our hope that this may lead to provable solutions to the problem of hallucinations in model-based MRI reconstruction.

\begin{credits}

\subsubsection{\ackname}
Suna Bu\u{g}day was supported by an Erasmus+ mobility grant during her stay at Ghent University.


\end{credits}

\bibliographystyle{splncs04}
\bibliography{main}

@article{gottschling2025troublesome,
  title={The Troublesome Kernel: On Hallucinations, No Free Lunches, and the Accuracy-Stability Tradeoff in Inverse Problems},
  author={Gottschling, Nina M and Antun, Vegard and Hansen, Anders C and Adcock, Ben},
  journal={SIAM Review},
  volume={67},
  number={1},
  pages={73--104},
  year={2025},
  publisher={SIAM}
}

@article{bellon1986mr,
  title={{MR artifacts: a review}},
  author={Bellon, Errol M and Haacke, E Mark and Coleman, Paul E and Sacco, Damon C and Steiger, David A and Gangarosa, Raymond E},
  journal={American Journal of Roentgenology},
  volume={147},
  number={6},
  pages={1271--1281},
  year={1986},
  publisher={American Roentgen Ray Society}
}

@article{zhang2022use,
  title={The use of deep learning methods in low-dose computed tomography image reconstruction: a systematic review},
  author={Zhang, Minghan and Gu, Sai and Shi, Yuhui},
  journal={Complex \& intelligent systems},
  volume={8},
  number={6},
  pages={5545--5561},
  year={2022},
  publisher={Springer}
}

@article{sidky2020cnns,
  title={{Do CNNs solve the CT inverse problem?}},
  author={Sidky, Emil Y and Lorente, Iris and Brankov, Jovan G and Pan, Xiaochuan},
  journal={IEEE Transactions on Biomedical Engineering},
  volume={68},
  number={6},
  pages={1799--1810},
  year={2020},
  publisher={IEEE}
}

@article{beylkin1987discrete,
  title={{Discrete Radon transform}},
  author={Beylkin, Gregory},
  journal={IEEE transactions on acoustics, speech, and signal processing},
  volume={35},
  number={2},
  pages={162--172},
  year={1987},
  publisher={IEEE}
}

@article{winograd1978computing,
  title={On computing the discrete {Fourier} transform},
  author={Winograd, Shmuel},
  journal={Mathematics of computation},
  volume={32},
  number={141},
  pages={175--199},
  year={1978}
}

@article{muckley2021results,
  title={{Results of the 2020 fastMRI challenge for machine learning MR image reconstruction}},
  author={Muckley, Matthew J and Riemenschneider, Bruno and Radmanesh, Alireza and Kim, Sunwoo and Jeong, Geunu and Ko, Jingyu and Jun, Yohan and Shin, Hyungseob and Hwang, Dosik and Mostapha, Mahmoud and others},
  journal={IEEE transactions on medical imaging},
  volume={40},
  number={9},
  pages={2306--2317},
  year={2021},
  publisher={IEEE}
}

@article{roemer1990nmr,
  title={The {NMR} phased array},
  author={Roemer, Peter B and Edelstein, William A and Hayes, Cecil E and Souza, Steven P and Mueller, Otward M},
  journal={Magnetic resonance in medicine},
  volume={16},
  number={2},
  pages={192--225},
  year={1990},
  publisher={Wiley Online Library}
}

@misc{ballinger2013acquisition,
  title = {Acquisition time},
  url = {http://dx.doi.org/10.53347/rID-22553},
  DOI = {10.53347/rid-22553},
  journal = {Radiopaedia.org},
  publisher = {Radiopaedia.org},
  author = {Ballinger,  J. and Wilczek,  Mateusz and Knipe,  Henry},
  year = {2013},
  month = apr 
}

@article{tropp2010computational,
  title={Computational methods for sparse solution of linear inverse problems},
  author={Tropp, Joel A and Wright, Stephen J},
  journal={Proceedings of the IEEE},
  volume={98},
  number={6},
  pages={948--958},
  year={2010},
  publisher={IEEE}
}

@article{xu2024hallucination,
  publtype={informal},
  author={Ziwei Xu and Sanjay Jain and Mohan S. Kankanhalli},
  title={Hallucination is Inevitable: An Innate Limitation of Large Language Models},
  year={2024},
  cdate={1704067200000},
  journal={CoRR},
  volume={abs/2401.11817},
  url={https://doi.org/10.48550/arXiv.2401.11817}
}

@article{farquhar2024detecting,
  title={Detecting hallucinations in large language models using semantic entropy},
  author={Farquhar, Sebastian and Kossen, Jannik and Kuhn, Lorenz and Gal, Yarin},
  journal={Nature},
  volume={630},
  number={8017},
  pages={625--630},
  year={2024},
  publisher={Nature Publishing Group UK London}
}

@article{ribes2008linear,
  title={Linear inverse problems in imaging},
  author={Ribes, Alejandro and Schmitt, Francis},
  journal={IEEE Signal Processing Magazine},
  volume={25},
  number={4},
  pages={84--99},
  year={2008},
  publisher={IEEE}
}

@article{muckley2020state,
  title={State-of-the-art machine learning {MRI} reconstruction in 2020: Results of the second {fastMRI} challenge},
  author={Muckley, Matthew J and Riemenschneider, Bruno and Radmanesh, Alireza and Kim, Sunwoo and Jeong, Geunu and Ko, Jingyu and Jun, Yohan and Shin, Hyungseob and Hwang, Dosik and Mostapha, Mahmoud and others},
  journal={arXiv preprint arXiv:2012.06318},
  volume={2},
  number={6},
  year={2020}
}

@article{aggarwal2018modl,
  title={{MoDL: Model-based deep learning architecture for inverse problems}},
  author={Aggarwal, Hemant K and Mani, Merry P and Jacob, Mathews},
  journal={IEEE transactions on medical imaging},
  volume={38},
  number={2},
  pages={394--405},
  year={2018},
  publisher={IEEE}
}

@article{szegedy2013intriguing,
  title={Intriguing properties of neural networks},
  author={Szegedy, Christian and Zaremba, Wojciech and Sutskever, Ilya and Bruna, Joan and Erhan, Dumitru and Goodfellow, Ian and Fergus, Rob},
  journal={arXiv preprint arXiv:1312.6199},
  year={2013}
}

@inproceedings{biggio2018wild,
  title={Wild patterns: Ten years after the rise of adversarial machine learning},
  author={Biggio, Battista and Roli, Fabio},
  booktitle={Proceedings of the 2018 ACM SIGSAC Conference on Computer and Communications Security},
  pages={2154--2156},
  year={2018}
}

@article{peck2023introduction,
  title={An introduction to adversarially robust deep learning},
  author={Peck, Jonathan and Goossens, Bart and Saeys, Yvan},
  journal={IEEE Transactions on Pattern Analysis and Machine Intelligence},
  year={2023},
  publisher={IEEE}
}

@inproceedings{chen1994basis,
  title={Basis pursuit},
  author={Chen, Shaobing and Donoho, David},
  booktitle={Proceedings of 1994 28th Asilomar Conference on Signals, Systems and Computers},
  volume={1},
  pages={41--44},
  year={1994},
  organization={IEEE}
}

@article{hammernik2018learning,
  title={Learning a variational network for reconstruction of accelerated {MRI} data},
  author={Hammernik, Kerstin and Klatzer, Teresa and Kobler, Erich and Recht, Michael P and Sodickson, Daniel K and Pock, Thomas and Knoll, Florian},
  journal={Magnetic resonance in medicine},
  volume={79},
  number={6},
  pages={3055--3071},
  year={2018},
  publisher={Wiley Online Library}
}

@article{wang2004image,
  title={Image quality assessment: from error visibility to structural similarity},
  author={Wang, Zhou and Bovik, Alan C and Sheikh, Hamid R and Simoncelli, Eero P},
  journal={IEEE transactions on image processing},
  volume={13},
  number={4},
  pages={600--612},
  year={2004},
  publisher={IEEE}
}

@article{mao2023connecting,
  title={Connecting certified and adversarial training},
  author={Mao, Yuhao and M{\"u}ller, Mark and Fischer, Marc and Vechev, Martin},
  journal={Advances in Neural Information Processing Systems},
  volume={36},
  pages={73422--73440},
  year={2023}
}

@inproceedings{lee2021towards,
 author = {Lee, Sungyoon and Lee, Woojin and Park, Jinseong and Lee, Jaewook},
 booktitle = {Advances in Neural Information Processing Systems},
 editor = {M. Ranzato and A. Beygelzimer and Y. Dauphin and P.S. Liang and J. Wortman Vaughan},
 pages = {953--964},
 publisher = {Curran Associates, Inc.},
 title = {Towards Better Understanding of Training Certifiably Robust Models against Adversarial Examples},
 url = {https://proceedings.neurips.cc/paper_files/paper/2021/file/07c5807d0d927dcd0980f86024e5208b-Paper.pdf},
 volume = {34},
 year = {2021}
}

@inproceedings{carlini2017adversarial,
  title={Adversarial examples are not easily detected: Bypassing ten detection methods},
  author={Carlini, Nicholas and Wagner, David},
  booktitle={Proceedings of the 10th ACM workshop on artificial intelligence and security},
  pages={3--14},
  year={2017}
}

@inproceedings{athalye2018obfuscated,
  title={Obfuscated gradients give a false sense of security: Circumventing defenses to adversarial examples},
  author={Athalye, Anish and Carlini, Nicholas and Wagner, David},
  booktitle={International conference on machine learning},
  pages={274--283},
  year={2018},
  organization={PMLR}
}

@article{tramer2020adaptive,
  title={On adaptive attacks to adversarial example defenses},
  author={Tramer, Florian and Carlini, Nicholas and Brendel, Wieland and Madry, Aleksander},
  journal={Advances in neural information processing systems},
  volume={33},
  pages={1633--1645},
  year={2020}
}

@inproceedings{biggio2013evasion,
  title={Evasion attacks against machine learning at test time},
  author={Biggio, Battista and Corona, Igino and Maiorca, Davide and Nelson, Blaine and {\v{S}}rndi{\'c}, Nedim and Laskov, Pavel and Giacinto, Giorgio and Roli, Fabio},
  booktitle={Joint European conference on machine learning and knowledge discovery in databases},
  pages={387--402},
  year={2013},
  organization={Springer}
}

@article{kaviani2022adversarial,
  title={Adversarial attacks and defenses on AI in medical imaging informatics: A survey},
  author={Kaviani, Sara and Han, Ki Jin and Sohn, Insoo},
  journal={Expert Systems with Applications},
  volume={198},
  pages={116815},
  year={2022},
  publisher={Elsevier}
}

@article{dong2024survey,
  title={Survey on adversarial attack and defense for medical image analysis: Methods and challenges},
  author={Dong, Junhao and Chen, Junxi and Xie, Xiaohua and Lai, Jianhuang and Chen, Hao},
  journal={ACM Computing Surveys},
  volume={57},
  number={3},
  pages={1--38},
  year={2024},
  publisher={ACM New York, NY}
}

@incollection{kurakin2018adversarial,
  title={Adversarial examples in the physical world},
  author={Kurakin, Alexey and Goodfellow, Ian J and Bengio, Samy},
  booktitle={Artificial intelligence safety and security},
  pages={99--112},
  year={2018},
  publisher={Chapman and Hall/CRC}
}

@article{costa2024deep,
  title={How deep learning sees the world: A survey on adversarial attacks \& defenses},
  author={Costa, Joana C and Roxo, Tiago and Proen{\c{c}}a, Hugo and Inacio, Pedro Ricardo Morais},
  journal={IEEE Access},
  volume={12},
  pages={61113--61136},
  year={2024},
  publisher={IEEE}
}

@article{del2023stability,
  title={Stability of image-reconstruction algorithms},
  author={del Aguila Pla, Pol and Neumayer, Sebastian and Unser, Michael},
  journal={IEEE Transactions on Computational Imaging},
  volume={9},
  pages={1--12},
  year={2023},
  publisher={IEEE}
}

@article{colbrook2022difficulty,
  title={{The difficulty of computing stable and accurate neural networks: On the barriers of deep learning and Smale’s 18th problem}},
  author={Colbrook, Matthew J and Antun, Vegard and Hansen, Anders C},
  journal={Proceedings of the National Academy of Sciences},
  volume={119},
  number={12},
  pages={e2107151119},
  year={2022},
  publisher={National Academy of Sciences}
}

@inproceedings{ronneberger2015unet,
  title={U-net: Convolutional networks for biomedical image segmentation},
  author={Ronneberger, Olaf and Fischer, Philipp and Brox, Thomas},
  booktitle={International Conference on Medical image computing and computer-assisted intervention},
  pages={234--241},
  year={2015},
  organization={Springer}
}

@article{genzel2022solving,
  title={Solving inverse problems with deep neural networks -- robustness included?},
  author={Genzel, Martin and Macdonald, Jan and M{\"a}rz, Maximilian},
  journal={IEEE transactions on pattern analysis and machine intelligence},
  volume={45},
  number={1},
  pages={1119--1134},
  year={2022},
  publisher={IEEE}
}

@inproceedings{darestani2021measuring,
  title={Measuring robustness in deep learning based compressive sensing},
  author={Darestani, Mohammad Zalbagi and Chaudhari, Akshay S and Heckel, Reinhard},
  booktitle={International Conference on Machine Learning},
  pages={2433--2444},
  year={2021},
  organization={PMLR}
}

@InProceedings{cheng2020addressing,
  title = 	 {{Addressing The False Negative Problem of Deep Learning MRI Reconstruction Models by Adversarial Attacks and Robust Training}},
  author =       {Cheng, Kaiyang and Caliv\'a, Francesco and Shah, Rutwik and Han, Misung and Majumdar, Sharmila and Pedoia, Valentina},
  booktitle = 	 {Proceedings of the Third Conference on Medical Imaging with Deep Learning},
  pages = 	 {121--135},
  year = 	 {2020},
  editor = 	 {Arbel, Tal and Ben Ayed, Ismail and de Bruijne, Marleen and Descoteaux, Maxime and Lombaert, Herve and Pal, Christopher},
  volume = 	 {121},
  series = 	 {Proceedings of Machine Learning Research},
  month = 	 {06--08 Jul},
  publisher =    {PMLR},
  url = 	 {https://proceedings.mlr.press/v121/cheng20a.html}
}

@article{uecker2014espirit,
  title={{ESPIRiT—an eigenvalue approach to autocalibrating parallel MRI: where SENSE meets GRAPPA}},
  author={Uecker, Martin and Lai, Peng and Murphy, Mark J and Virtue, Patrick and Elad, Michael and Pauly, John M and Vasanawala, Shreyas S and Lustig, Michael},
  journal={Magnetic resonance in medicine},
  volume={71},
  number={3},
  pages={990--1001},
  year={2014},
  publisher={Wiley Online Library}
}

@inproceedings{cohen2019certified,
  title={Certified adversarial robustness via randomized smoothing},
  author={Cohen, Jeremy and Rosenfeld, Elan and Kolter, Zico},
  booktitle={international conference on machine learning},
  pages={1310--1320},
  year={2019},
  organization={PMLR}
}

@article{peck2024robust,
  title={Robust width: A lightweight and certifiable adversarial defense},
  author={Peck, Jonathan and Goossens, Bart},
  journal={arXiv preprint arXiv:2405.15971},
  year={2024}
}

@inproceedings{chen2013robust,
  title={Robust sparse regression under adversarial corruption},
  author={Chen, Yudong and Caramanis, Constantine and Mannor, Shie},
  booktitle={International conference on machine learning},
  pages={774--782},
  year={2013},
  organization={PMLR}
}

@inproceedings{chattopadhyay2019curse,
  title={Curse of dimensionality in adversarial examples},
  author={Chattopadhyay, Nandish and Chattopadhyay, Anupam and Gupta, Sourav Sen and Kasper, Michael},
  booktitle={2019 International Joint Conference on Neural Networks (IJCNN)},
  pages={1--8},
  year={2019},
  organization={IEEE}
}

@misc{gilmer2018adversarial,
title={Adversarial Spheres},
author={Justin Gilmer and Luke Metz and Fartash Faghri and Sam Schoenholz and Maithra Raghu and Martin Wattenberg and Ian Goodfellow},
year={2018},
url={https://openreview.net/forum?id=SyUkxxZ0b},
}

@article{sulam2020adversarial,
  title={Adversarial robustness of supervised sparse coding},
  author={Sulam, Jeremias and Muthukumar, Ramchandran and Arora, Raman},
  journal={Advances in neural information processing systems},
  volume={33},
  pages={2110--2121},
  year={2020}
}

@article{salman2020denoised,
  title={Denoised smoothing: A provable defense for pretrained classifiers},
  author={Salman, Hadi and Sun, Mingjie and Yang, Greg and Kapoor, Ashish and Kolter, J Zico},
  journal={Advances in Neural Information Processing Systems},
  volume={33},
  pages={21945--21957},
  year={2020}
}

@article{heckel2024deep,
  title={Deep Learning for Accelerated and Robust {MRI} Reconstruction: a Review},
  author={Heckel, Reinhard and Jacob, Mathews and Chaudhari, Akshay and Perlman, Or and Shimron, Efrat},
  journal={arXiv preprint arXiv:2404.15692},
  year={2024}
}

@book{foucart2013,
  title = {A Mathematical Introduction to Compressive Sensing},
  ISBN = {9780817649487},
  ISSN = {2296-5017},
  url = {http://dx.doi.org/10.1007/978-0-8176-4948-7},
  DOI = {10.1007/978-0-8176-4948-7},
  journal = {Applied and Numerical Harmonic Analysis},
  publisher = {Springer New York},
  author = {Foucart,  Simon and Rauhut,  Holger},
  year = {2013}
}

@inproceedings{morshuis2022adversarial,
  title={Adversarial robustness of {MR} image reconstruction under realistic perturbations},
  author={Morshuis, Jan Nikolas and Gatidis, Sergios and Hein, Matthias and Baumgartner, Christian F},
  booktitle={International Workshop on Machine Learning for Medical Image Reconstruction},
  pages={24--33},
  year={2022},
  organization={Springer}
}

@article{bhadra2021hallucinations,
  title={On hallucinations in tomographic image reconstruction},
  author={Bhadra, Sayantan and Kelkar, Varun A and Brooks, Frank J and Anastasio, Mark A},
  journal={IEEE transactions on medical imaging},
  volume={40},
  number={11},
  pages={3249--3260},
  year={2021},
  publisher={IEEE}
}

@article{chambolle2010introduction,
  title={An introduction to total variation for image analysis},
  author={Chambolle, Antonin and Caselles, Vicent and Cremers, Daniel and Novaga, Matteo and Pock, Thomas and others},
  journal={Theoretical foundations and numerical methods for sparse recovery},
  volume={9},
  number={263-340},
  pages={227},
  year={2010}
}

@article{torrence1998practical,
  title={A practical guide to wavelet analysis},
  author={Torrence, Christopher and Compo, Gilbert P},
  journal={Bulletin of the American Meteorological society},
  volume={79},
  number={1},
  pages={61--78},
  year={1998},
  publisher={American Meteorological Society}
}

@article{block2007undersampled,
  title={{Undersampled radial MRI with multiple coils. Iterative image reconstruction using a total variation constraint}},
  author={Block, Kai Tobias and Uecker, Martin and Frahm, Jens},
  journal={Magnetic Resonance in Medicine: An Official Journal of the International Society for Magnetic Resonance in Medicine},
  volume={57},
  number={6},
  pages={1086--1098},
  year={2007},
  publisher={Wiley Online Library}
}

@article{kawata2007constrained,
  title={Constrained iterative reconstruction by the conjugate gradient method},
  author={Kawata, Satoshi and Nalcioglu, Orhan},
  journal={IEEE transactions on medical imaging},
  volume={4},
  number={2},
  pages={65--71},
  year={2007},
  publisher={IEEE}
}

@article{beck2009fast,
  title={A fast iterative shrinkage-thresholding algorithm for linear inverse problems},
  author={Beck, Amir and Teboulle, Marc},
  journal={SIAM journal on imaging sciences},
  volume={2},
  number={1},
  pages={183--202},
  year={2009},
  publisher={SIAM}
}

@inproceedings{aromal2024fista,
  title={{FISTA-NET}: Compressed Sensing {MRI} Reconstruction Using Unrolled Iterative Networks},
  author={Aromal, CJ and Datta, Sumit},
  booktitle={2024 IEEE 21st India Council International Conference (INDICON)},
  pages={1--6},
  year={2024},
  organization={IEEE}
}

@InProceedings{sriram2020end,
author="Sriram, Anuroop
and Zbontar, Jure
and Murrell, Tullie
and Defazio, Aaron
and Zitnick, C. Lawrence
and Yakubova, Nafissa
and Knoll, Florian
and Johnson, Patricia",
editor="Martel, Anne L.
and Abolmaesumi, Purang
and Stoyanov, Danail
and Mateus, Diana
and Zuluaga, Maria A.
and Zhou, S. Kevin
and Racoceanu, Daniel
and Joskowicz, Leo",
title="End-to-End Variational Networks for Accelerated MRI Reconstruction",
booktitle="Medical Image Computing and Computer Assisted Intervention -- MICCAI 2020",
year="2020",
publisher="Springer International Publishing",
address="Cham",
pages="64--73",
isbn="978-3-030-59713-9"
}

@article{zbontar2018fastmri,
  title={{fastMRI: An open dataset and benchmarks for accelerated MRI}},
  author={Zbontar, Jure and Knoll, Florian and Sriram, Anuroop and Murrell, Tullie and Huang, Zhengnan and Muckley, Matthew J and Defazio, Aaron and Stern, Ruben and Johnson, Patricia and Bruno, Mary and others},
  journal={arXiv preprint arXiv:1811.08839},
  year={2018}
}

\clearpage

\appendix

\crefalias{section}{appendix}
\numberwithin{figure}{section}
\numberwithin{algorithm}{section}

\section{Pseudocode of the attack}\label{app:pseudocode}

\begin{algorithm}
    \caption{Masked Iterative FGSM Attack}\label{alg:attack}
    \begin{description}
      \item[$F$] Reconstruction map.
      \item[$\bm z$] Raw $k$-space input sample.
      \item[$\bm y_t$] Target reconstruction with artificial detail.
      \item[$\bm m$] Binary mask delineating the target region.
      \item[$\varepsilon$] Maximum perturbation magnitude.
      \item[$\alpha$] Step size per iteration.
      \item[$T$] Number of iterations.
    \end{description}
    \begin{algorithmic}[1]
    \Function{MaskedIterativeFGSM}{$F$, $\bm z$, $\bm y_t$, $\bm m$, $\varepsilon$, $\alpha$, $T$}
        \State $\bm y_0 \gets F(\bm z)$ \Comment{Obtain clean reconstruction}
        \State $\bm\delta \sim \mathrm{Unif}(-\varepsilon, \varepsilon)$ \Comment{Random initial perturbation}
        \State $\bm\delta^\star \gets \bm\delta$ \Comment{Initial best perturbation}
        \State $L_{\min} \gets +\infty$ \Comment{Initial best loss value}
        
        \LComment{Optimization loop}
        \For{$t \gets 1$ \textbf{to} $T$}
            \LComment{Compute perturbed sample}
            \State $\bm z_{\text{adv}} \gets \operatorname{clip}(\Re(\bm z) + \bm\delta, 0, 1) + \Im(\bm z)$

            \LComment{Compute loss}
            \State $L_1 \gets \|\bm m \odot (F(\bm z_{\text{adv}}) - \bm y_t)\|_2^2$
            \State $L_2 \gets \|(1 - \bm m) \odot (F(\bm z_{\text{adv}}) - F(\bm z))\|_2^2$
            \State $L \gets \dfrac{L_1}{\norm{\bm m}_1} + \dfrac{L_2}{\norm{1 - \bm m}_1}$

            \LComment{Keep track of the best perturbation}
            \If{$L < L_{\min}$}
                \State $\bm\delta^\star \gets \bm\delta$
                \State $L_{\min} \gets L$
            \EndIf

            \LComment{Compute update}
            \State $\bm\delta \gets \bm\delta - \alpha \cdot \text{sign}(\nabla_{\bm\delta} L)$
            \State $\bm\delta \gets \operatorname{clip}(\bm\delta, -\varepsilon, \varepsilon)$
        \EndFor

        \State \Return $\bm\delta^\star$
    \EndFunction
    \end{algorithmic}
\end{algorithm}

\clearpage

\section{Qualitative assessment of the attack}\label{app:qualitative}
\begin{figure}
    \begin{center}
        \subfloat[][sc-knee\label{fig:unet-best1}]{
            \begin{tabular}{@{} c|c @{}}
                \includegraphics[width=.4\columnwidth]{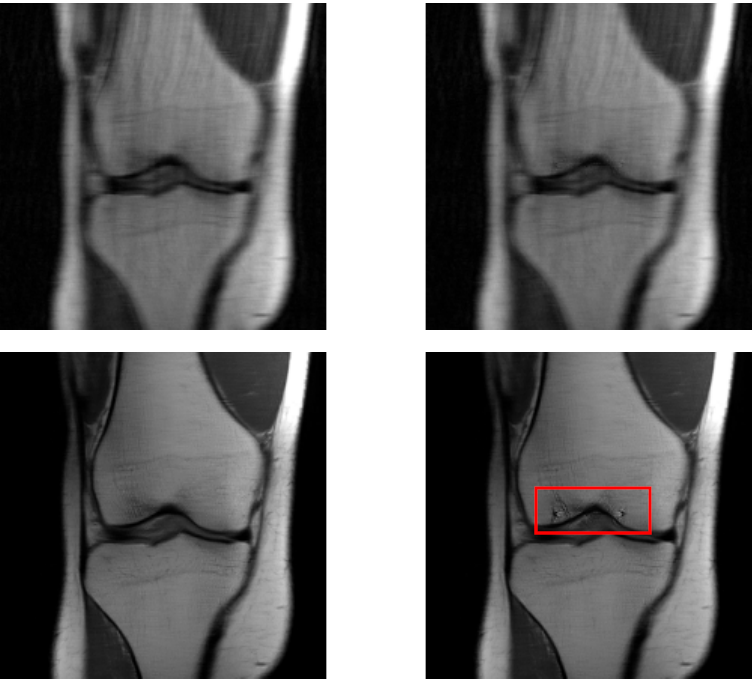}
                &
                \includegraphics[width=.4\columnwidth]{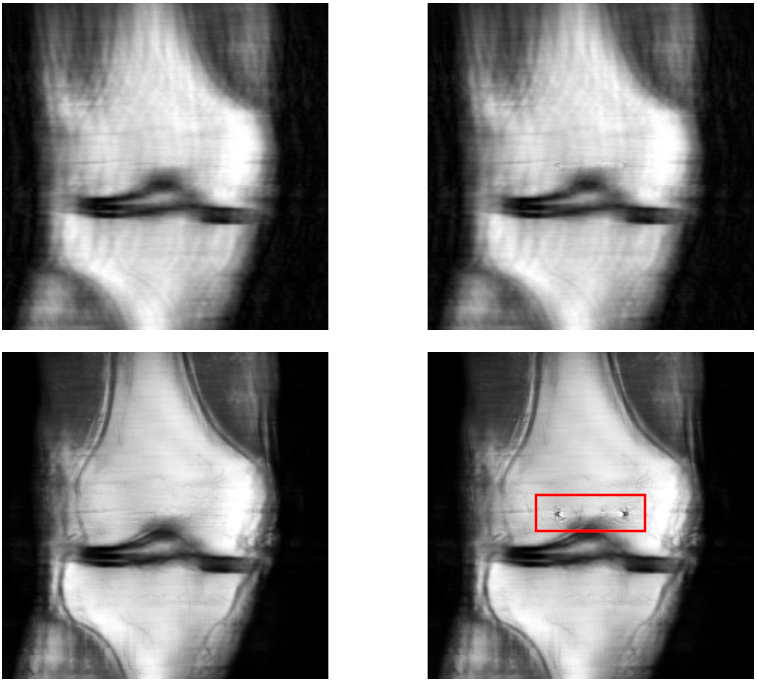}
            \end{tabular}
        }
        
        \subfloat[][mc-knee\label{fig:unet-best2}]{
            \begin{tabular}{@{} c|c @{}}
                \includegraphics[width=.4\columnwidth]{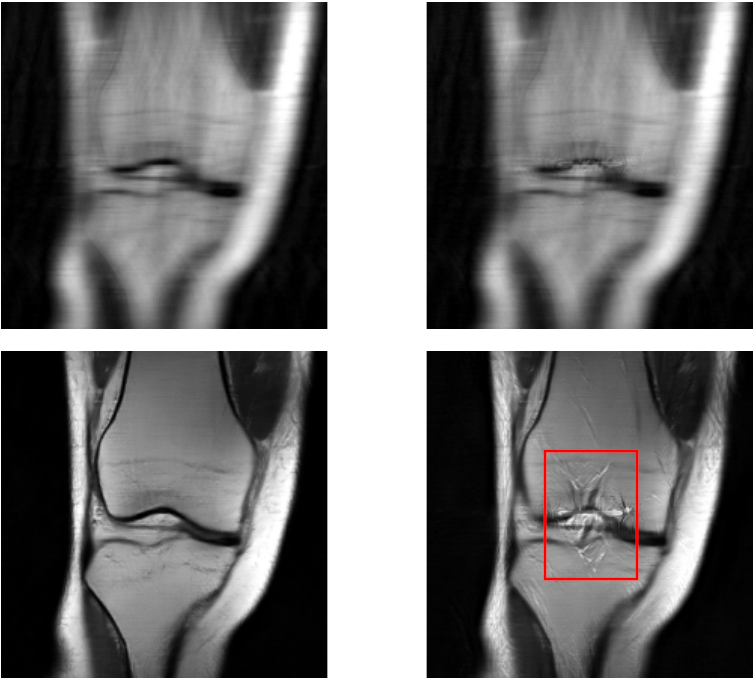}
                &
                \includegraphics[width=.4\columnwidth]{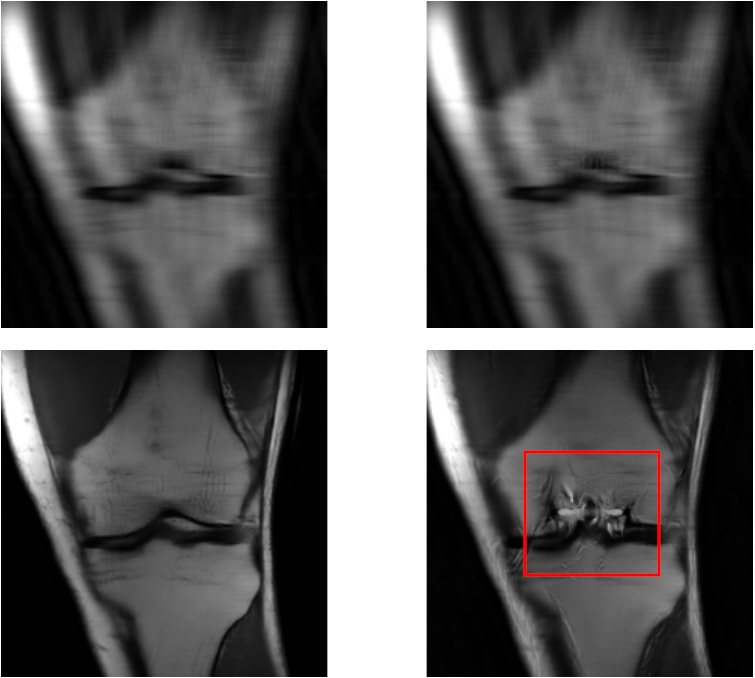}
            \end{tabular}
        }
        
        \subfloat[][mc-brain\label{fig:unet-best3}]{
            \begin{tabular}{@{} c|c @{}}
                \includegraphics[width=.4\columnwidth]{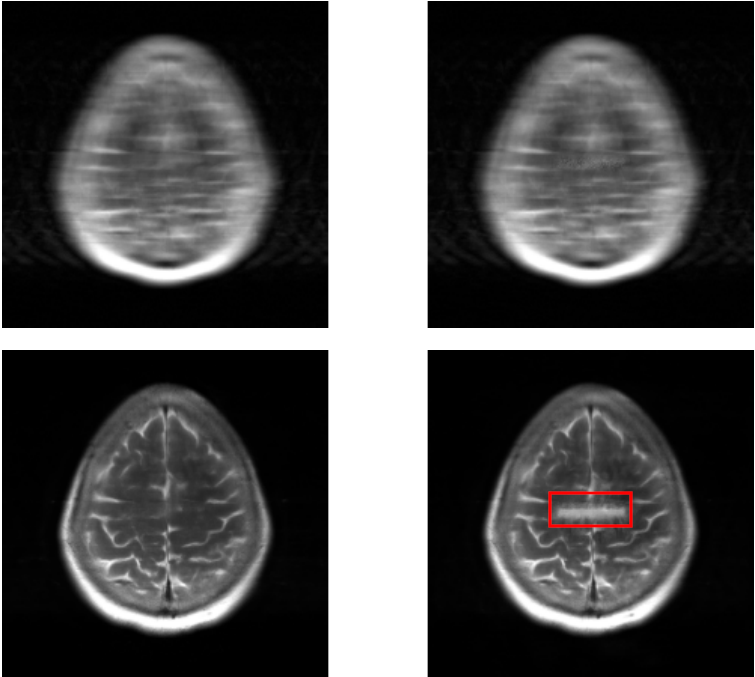}
                &
                \includegraphics[width=.4\columnwidth]{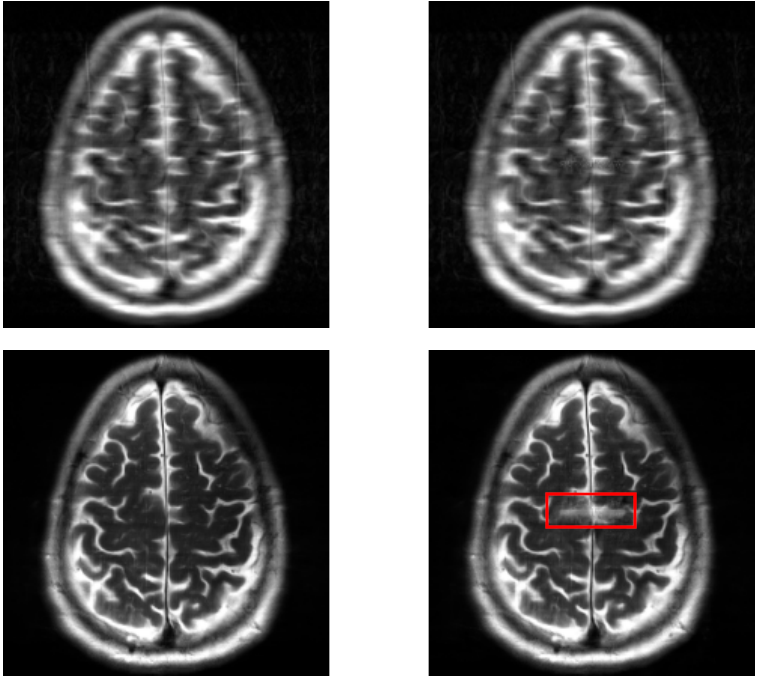}
            \end{tabular}
        }
    \end{center}
    \caption{Examples of successful attacks on the UNet model.
    For each data set, the first row displays the input samples (original on the left, adversarially perturbed on the right) and the second row displays the corresponding outputs.
    The generated hallucinations are annotated in red.}
    \label{fig:unet-best}
\end{figure}

\begin{figure}
    \begin{center}
        \subfloat[][mc-knee\label{fig:varnet-best1}]{
            \begin{tabular}{@{} c|c @{}}
                \includegraphics[width=.4\columnwidth]{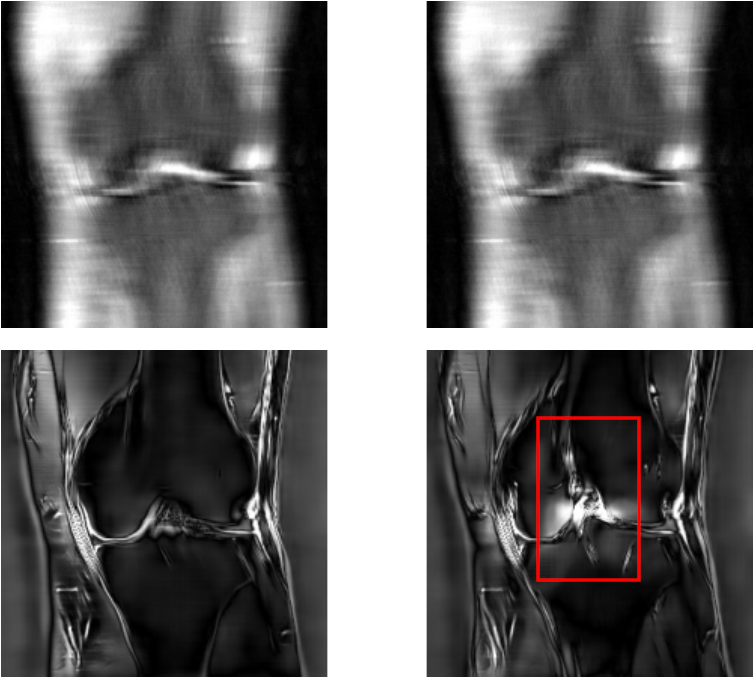}
                &
                \includegraphics[width=.4\columnwidth]{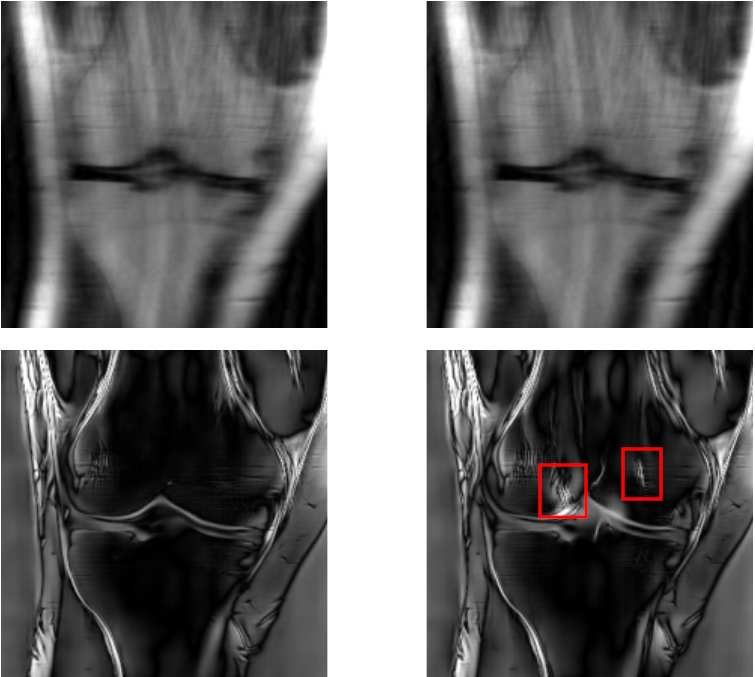}
            \end{tabular}
        }
        
        \subfloat[][mc-brain\label{fig:varnet-best2}]{
            \begin{tabular}{@{} c|c @{}}
                \includegraphics[width=.4\columnwidth]{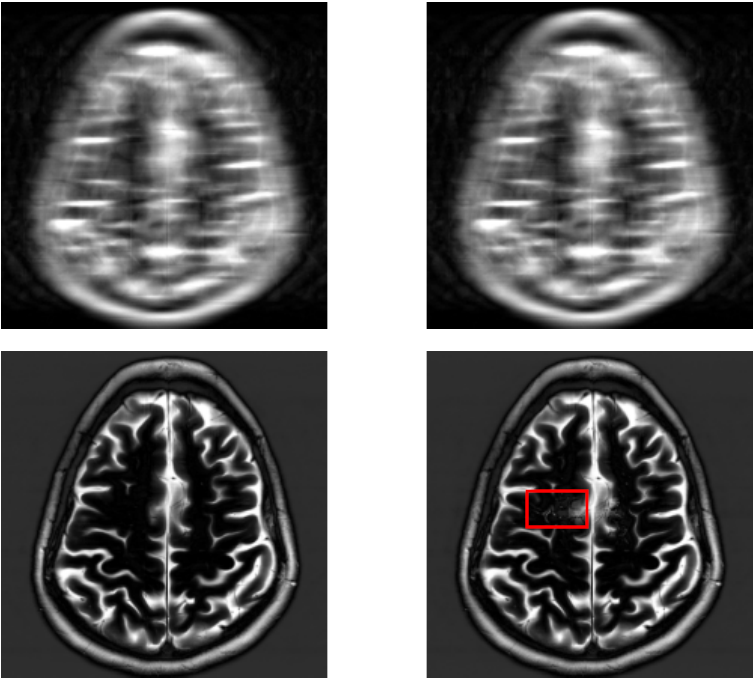}
                &
                \includegraphics[width=.4\columnwidth]{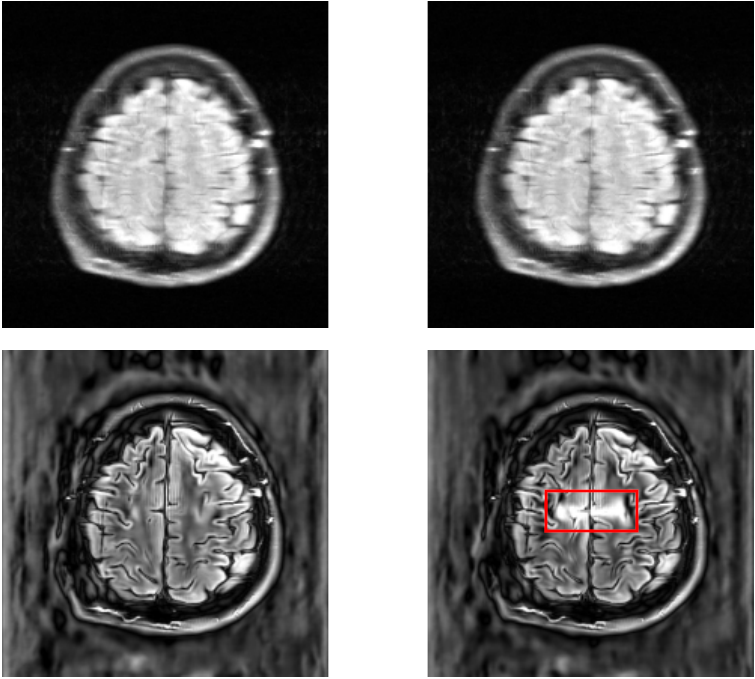}
            \end{tabular}
        }
    \end{center}
    \caption{Examples of successful attacks on the E2E-VarNet model.
    For each data set, the first row displays the input samples (original on the left, adversarially perturbed on the right) and the second row displays the corresponding outputs.
    The generated hallucinations are annotated in red.}
    \label{fig:varnet-best}
\end{figure}

For a qualitative assessment of the results, we show some examples in \cref{fig:unet-best} for the UNet model and \cref{fig:varnet-best} for the E2E-VarNet model of reconstructions where the attack was successful.
These figures are structured as 2x2 panels of images where the first row displays the original and perturbed input samples and the second row displays the corresponding model-based reconstructions.
Areas which we believe contain hallucinatory structures are highlighted in red.

We notice from \cref{fig:unet-best1,fig:unet-best2} that the multi-coil images seem to be more vulnerable than the single-coil ones, in the sense that the resulting distortions are more severe for multi-coil data.
The generated perturbations also seem easier to spot for single-coil data.
This may be explained by the fact that perturbations in the multi-coil images can be more ``spread out'' across the coils, whereas with single-coil data this is not possible.
This is consistent with the observation that vulnerability to adversarial examples increases with data dimensionality~\cite{chattopadhyay2019curse,gilmer2018adversarial}.
Although we cannot compare to single-coil data for the E2E-VarNet, we do notice large distortions in \cref{fig:varnet-best1} as well that go far beyond the boundaries of our inserted detail.
On the multi-coil brain data, the hallucinations seem less severe for both models compared to the knee data, but the distortions can still be significant as they tend to resemble non-existent sulci.
For knee images, the distortions appear to substantially change the structure of the knee joint, especially on multi-coil data.

We conclude that the adversarially perturbed samples can lead to realistic reconstructions which exhibit biologically plausible distortions that could mislead expert interpretation,
and that the insertion of the artificial detail will often cause further distortion beyond the original target region.
For both models, this is especially apparent in the multicoil knee data, where the shape of the knee joint tends to change significantly when the detail is inserted.

\end{document}